\pdfoutput=1
\documentclass[11pt]{article}

\usepackage[final]{acl}

\usepackage{times}
\usepackage{latexsym}

\usepackage[T1]{fontenc}
\usepackage[utf8]{inputenc}

\usepackage{microtype}

\definecolor{lightgreen}{RGB}{159,240,186}  % 使用RGB值
\definecolor{lightred}{RGB}{255,153,153}  % 使用RGB值
\usepackage{inconsolata}
\usepackage{graphicx}
\usepackage{booktabs}
\usepackage{bm}
\usepackage{colortbl}
\usepackage{balance}
\usepackage{booktabs} % For formal tables
\usepackage{graphics}
\usepackage{array}
\usepackage{flushend}
\usepackage[english]{babel}
\usepackage[T1]{fontenc}
\usepackage{textcomp}
\usepackage{latexsym}
\usepackage{graphicx}
\usepackage{multirow}
\usepackage[labelformat=simple]{subcaption}
\usepackage{algorithm}
\usepackage{algpseudocode}
\usepackage{amsmath}
\usepackage{tabularx}
\usepackage{pifont}
\usepackage{amsfonts}
\usepackage{enumitem}

\title{Learning Interpretable Legal Case Retrieval via Knowledge-Guided \\ Case Reformulation}

\author{Chenlong Deng$^{1}$, Kelong Mao$^{1}$, Zhicheng Dou$^{1}$\thanks{Corresponding author.}\\ 
    $^1$Gaoling School of Artificial Intelligence, Renmin University of China \\ 
    \texttt{\{dengchenlong,dou\}@ruc.edu.cn} \\
}

\begin{document}
\maketitle
\begin{abstract}
Legal case retrieval for sourcing similar cases is critical in upholding judicial fairness.
Different from general web search, legal case retrieval involves processing lengthy, complex, and highly specialized legal documents. Existing methods in this domain often overlook the incorporation of legal expert knowledge, which is crucial for accurately understanding and modeling legal cases, leading to unsatisfactory retrieval performance. 
This paper introduces KELLER, a legal knowledge-guided case reformulation approach based on large language models (LLMs) for effective and interpretable legal case retrieval.
By incorporating professional legal knowledge about crimes and law articles, we enable large language models to accurately reformulate the original legal case into concise sub-facts of crimes, which contain the essential information of the case.
% We then employ an efficient cross-matching module that measures the relevance of a candidate document case concerning a given query case based on their extracted sub-facts.
% We further design a dual-level contrastive learning method to facilitate the modeling of the interactions among these sub-facts for improved retrieval.
Extensive experiments on two legal case retrieval benchmarks demonstrate superior retrieval performance and robustness on complex legal case queries of KELLER over existing methods.
% The significant improvement in the complex query subset further demonstrates the robustness of our proposed approach.
\end{abstract}

\section{Introduction}
% Legal case retrieval empowers legal experts to thoroughly analyze applicable precedents before making decisions, thereby supporting the principles of justice and fairness. It has become an indispensable part of the two major legal systems worldwide. In the civil law system, while adherence to past cases (i.e., ``stare decisis'') is not a binding obligation, referencing prior relevant cases remains a recommended practice for judges to enhance the accuracy and credibility of judgments. In recent years, the surge in digital case records has underscored the necessity for advanced legal case retrieval systems, leading to an integration of interdisciplinary research in law and computer science.
Legal case retrieval is vital for legal experts to make informed decisions by thoroughly analyzing relevant precedents, which upholds justice and fairness~\cite{ hamann2019german}. This practice is crucial in both common law and civil law systems globally~\cite{lastres2015rebooting, harris2002final}. In civil law, although following past cases (known as "stare decisis") is not mandatory, judges are still highly advised to consider previous cases to improve the accuracy and trustworthiness of their judgments. 

\begin{figure}[!t]
	\centering
	% \vspace{-0.05cm}
	% \setlength{\abovecaptionskip}{0.1cm}
	\includegraphics[width=0.94\linewidth]{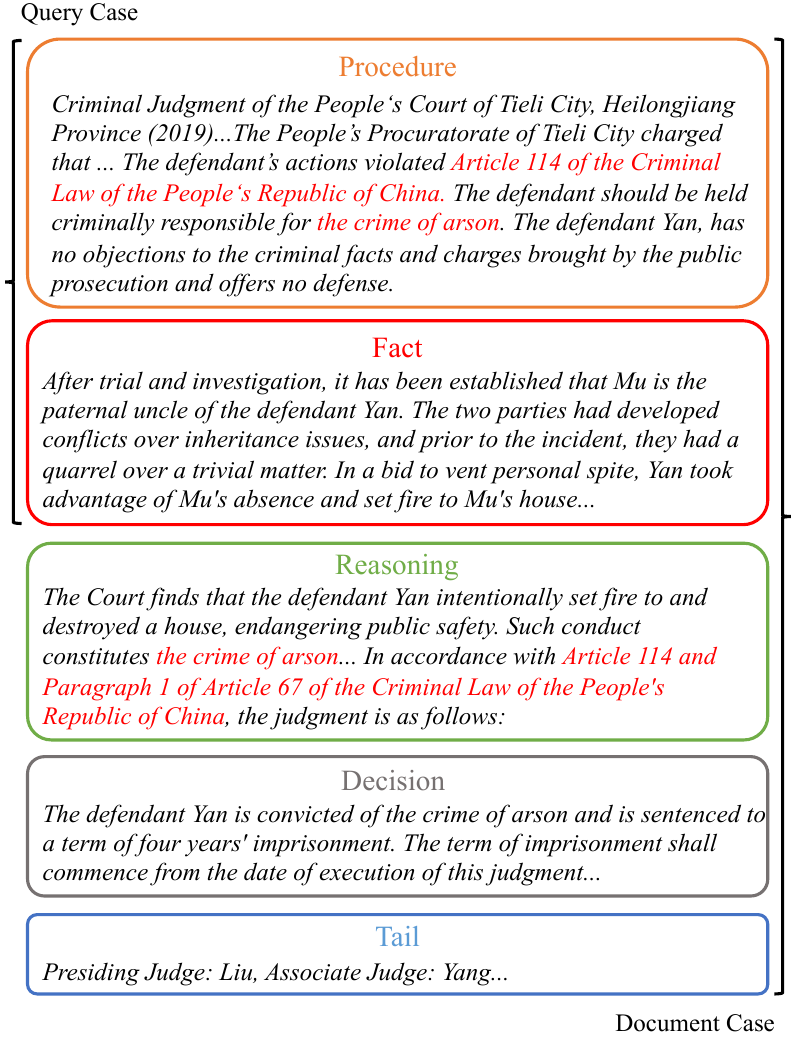}
 % \vspace{1ex}
	\caption{The query case and candidate document case examples. The query case typically contains only partial content since it has not been adjudicated. Extractable crimes and law articles are highlighted in red.}
	\label{fig:Fig_CaseStructure}
 % \vspace{-2ex}
\end{figure}

In legal case retrieval, both the query and the document are structured legal cases, distinguishing the task from other information retrieval (IR) tasks. Specifically, as shown in Figure~\ref{fig:Fig_CaseStructure}, a legal case document comprises several sections, such as procedure, facts, and the court's decision, making it much longer than typical queries and passages in the standard ad-hoc search tasks. Its average text length often exceeds the maximum input limits of popular retrievers, such as 512 tokens~\cite{bert}.
Moreover, a legal case may encompass multiple, distinct criminal behaviors. Comprehensively considering all criminal behaviors of a legal case is important in determining its matching relevance with a query case. However, these key criminal descriptions are usually dispersed throughout the lengthy contents, which can significantly affect the effectiveness of traditional long document modeling strategies like FirstP and MaxP~\cite{dai2019deeper} in the legal domain.

To tackle the challenge of comprehending long and complex legal cases, previous works mainly fall into two categories. The first approach focuses on expanding the context window size~\cite{xiao2021lawformer} or splitting legal cases into passages~\cite{shao2020bert}. However, given the specialized and complex nature of legal texts, merely increasing the context window size still proves insufficient for significantly improving the retrieval performance.
In contrast, the second approach performs direct text summarization~\cite{DBLP:conf/desires/AskariV21, DBLP:conf/adc/TangQL23} or embedding-level summarization~\cite{DBLP:conf/sigir/YuS0DCXW22} on the legal case, aiming to only keep the most crucial information for assessing the relevance between legal cases.
However, they typically only rely on heuristic rules or the models' inherent knowledge for summarization. As the legal domain is highly specialized, existing approaches that overlook professional legal knowledge (e.g., law articles) are likely to perform inaccurate summarization.
% For example, \citet{DBLP:conf/desires/AskariV21} adopt extractive or generative techniques to summarize cases. \citet{shao2020bert} hierarchically model paragraph-level interactions and finally aggregate matching signals between paragraphs into a score. \citet{DBLP:conf/sigir/YuS0DCXW22} utilize BERT to encode each sentence, transforming token-level inputs into sentence-level representations.
% With the advent of large language models (LLMs), \citet{DBLP:conf/adc/TangQL23} prompt LLMs to ``Summarize in 50 words'' to extract essential legal facts from cases. 

% In contrast, we posit that utilizing the specialized knowledge of authoritative legal theory would be beneficial to enhance the model’s comprehension of legal cases and improve retrieval effectiveness.

In this paper, we present a Knowledge-guidEd case reformuLation approach for LEgal case Retrieval, named KELLER.
Our main idea is to leverage professional legal knowledge to guide large language models (LLMs) to summarize the corresponding key sub-facts for the crimes of the legal cases, and then directly learn to model case relevance based on these crucial and concise sub-facts.
% Figure~\ref{} shows a high-level illustration of our approach. \textcolor{blue}{this figure should have concrete crime, sub-facts examples}.

\begin{figure*}[]
	\centering
	\includegraphics[width=0.95\linewidth]{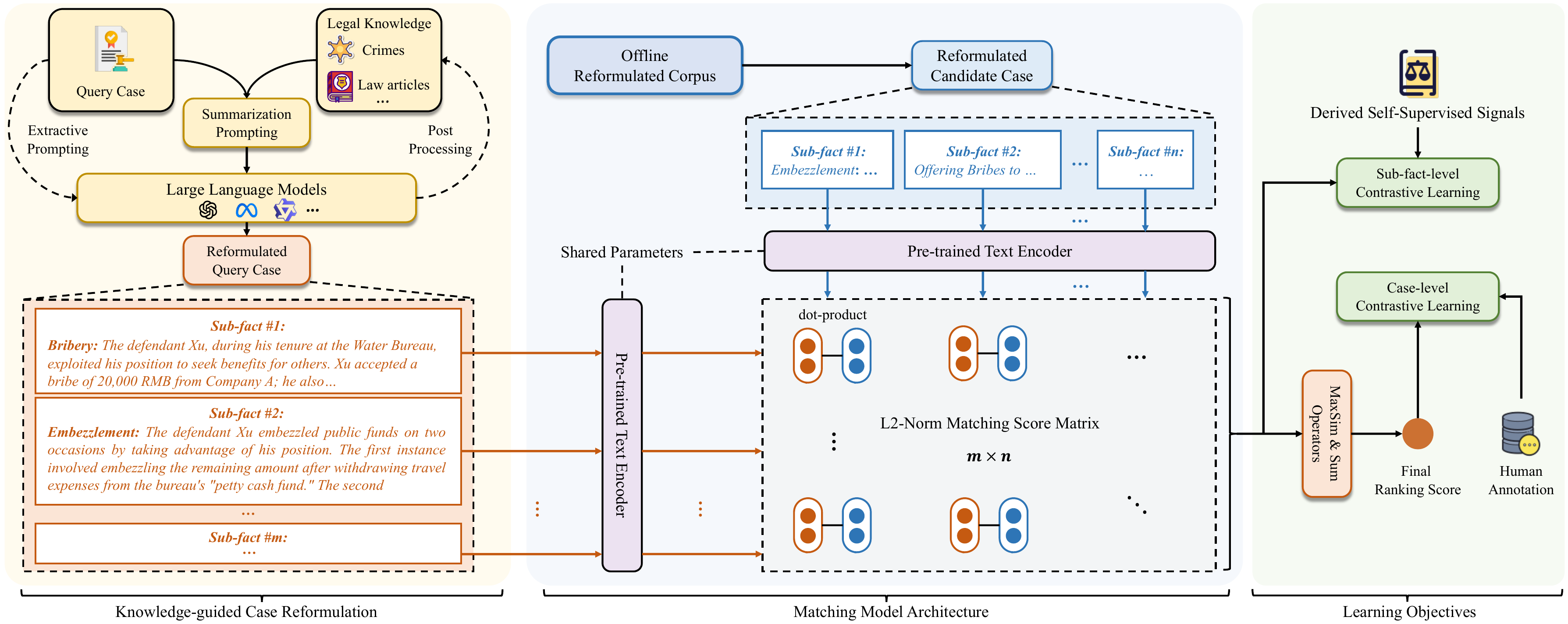}
	\caption{Overview of KELLER. We first perform legal knowledge-guided prompting to reformulate the legal cases into a series of crucial and concise sub-facts. Then, we directly model the case relevance based on the sub-facts. The model is trained at both the coarse-grained case level and the fine-grained sub-fact level via contrastive learning.}
	\label{fig:Fig_Model}
\end{figure*}

Due to the specialization and complexity of the legal case, it is quite challenging to directly summarize the corresponding key sub-facts for all the crimes from the legal case, even using advanced LLMs~\cite{ DBLP:conf/adc/TangQL23}. 
To address this problem, we propose a two-step legal knowledge-guided prompting method, as illustrated in the left side of Figure~\ref{fig:Fig_Model}. In the initial step, we prompt LLM to extract all of the crimes and law articles contained in the legal case and then perform post-processing on them to establish correct mappings between the crimes and law articles by referring to the legal expert database.
In the next step, we prompt LLM with the extracted ``crime-article '' pairs to summarize the sub-fact of the crime from the legal case. The intermediate law articles, serving as high-level abstractions of the actual criminal events, can largely reduce the difficulty of identifying the corresponding sub-fact for the crime and improve accuracy.
Figure~\ref{fig:Fig_CaseStudy_CaseReformulation} shows an example of three summarized sub-facts from a legal case.

Then, we directly model the case relevance based on these sub-facts because they are not only the most crucial information for relevance judgment in legal case retrieval but are also concise enough to meet the text length limitations of popular pre-trained retrieval models. For the comprehensive consideration of effectiveness, efficiency, and interoperability, we adopt the simple \textit{MaxSim} and \textit{Sum} operators to aggregate the relevance scores between query and document sub-facts to get the final case relevance score.
The model is trained with dual-level contrastive learning to comprehensively capture the matching signals at the case level and the sub-fact level.
On two widely-used datasets, we show that KELLER achieves new state-of-the-art results in both zero-shot and fine-tuning settings. Remarkably, KELLER demonstrates substantial improvements in handling complex queries.

Our main contributions can be summarized as:

(1) We propose to leverage professional legal knowledge about crimes and law articles to equip LLM with much-improved capabilities for summarizing essential sub-facts from complex cases.

(2) We suggest performing simple \textit{MaxSim} and \textit{Sum} aggregation directly on those refined sub-facts to achieve effective and interpretable legal retrieval.

(3) We introduce dual-level contrastive learning that enables the model to capture multi-granularity matching signals from both case-level and sub-fact-level for enhanced retrieval performance.

% \begin{figure*}[]
% 	\centering
% 	% \vspace{-0.05cm}
% 	% \setlength{\abovecaptionskip}{0.1cm}
% 	\includegraphics[width=\linewidth]{Figures/Fig_Idea.pdf}
%  % \vspace{1ex}
% 	\caption{A conceptual comparison of our approach and mainstream techniques in legal case retrieval. The \textcolor{red}{red} and \textcolor{blue}{blue} text in the case content represents two distinct sub-facts.}
% 	\label{fig:Fig_Idea}
%  % \vspace{-2ex}
% \end{figure*}

\section{Related Work}
\noindent \textbf{Legal case retrieval.}
Existing legal case retrieval methods are categorized into statistical and neural models. Statistical models, notably the BM25 algorithm, can be enhanced by incorporating legal expert knowledge such as legal summarization~\cite{tran2020encoded, DBLP:conf/desires/AskariV21}, issue elements~\cite{DBLP:conf/kes/ZengWZK05} and ontology~\cite{saravanan2009improving}. Neural models have been advanced through deep learning and the use of pre-trained language models~\cite{bert, zhong2019openclap, DBLP:journals/corr/abs-2010-02559, zhang2023cfgl}. Recent advancements in this domain include the design of specialized pre-training tasks tailored for legal case retrieval, which yields remarkable improvements in retrieval metrics~\cite{DBLP:conf/sigir/LiACDW0CT23, DBLP:conf/emnlp/MaWSA023}.

Due to the limitations of neural models in handling long texts, researchers mainly focus on processing lengthy legal documents by isolating the "fact description" section and truncating it to fit the model's input constraints~\cite{ma2021lecard, DBLP:conf/acl/YaoXWL0TLLSS22, DBLP:conf/emnlp/MaWSA023, DBLP:conf/sigir/LiACDW0CT23}.
To overcome the long-text problem, some other strategies include segmenting texts into paragraphs for interaction modeling~\cite{shao2020bert}, employing architectures like Longformer for extensive pre-training on legal texts~\cite{xiao2021lawformer}, and transforming token-level inputs into sentence-level encoding~\cite{DBLP:conf/sigir/YuS0DCXW22}. \\

\noindent \textbf{Query rewriting with LLMs.}
Recently, researchers naturally employ LLMs to enhance the effectiveness of query rewriting~\cite{LLM4IR_Survey, llm4cs, Query_Rewriting_for_Retrieval-Augmented_Large_Language_Models, Query2Doc, CoT_QueryRewriting}. For instance, HyDE~\cite{HyDE} creates pseudo passages for better query answers, integrating them into a vector for retrieval, while Query2Doc~\cite{Query2Doc} employs few-shot methods to generate precise responses. Furthermore, \citet{CoT_QueryRewriting} explores LLMs' reasoning capacities to develop "Chain-of-Thoughts" responses for complex queries.
However, the above methods struggle with legal case retrieval, where both queries and documents are lengthy cases. In the legal domain, PromptCase~\cite{DBLP:conf/adc/TangQL23} attempts to address this by summarizing case facts within 50 words, but this approach often misses important details as many cases feature multiple independent facts. 

% Additionally, most advanced LLMs are not specialized in legal texts, which can lead to critical omissions in summaries. To enhance content quality, we suggest using expert knowledge to help LLMs better reformulate complex legal cases.

% Recently, large language models (LLMs) have demonstrated remarkable proficiency across various IR tasks~\cite{DBLP:journals/corr/abs-2308-07107}. \citet{DBLP:conf/adc/TangQL23} prompt LLMs ``Summarize in 50 words'' to extract essential legal facts from cases. Nonetheless, summaries generated by guidance-free language models run the risk of omitting critical information. We propose the utilization of expert knowledge to guide LLMs in reformulating cases into structured legal content. This strategy is more trustworthy for expert users and ensures the maximal preservation of key content.

\section{Methodology}
% KELLER consists of three main components: legal knowledge-guided case reformulation, relevance modeling, and dual-level contrastive learning.
% The overview is shown in Figure~\ref{fig:Fig_Model}. 

% Generally, we first employ the LLM to generate a set of refined sub-facts from the intricate legal case texts by considering professional legal knowledge, such as crimes and law articles. Then, we directly model the relevance of these sub-facts to enhance retrieval performance.

In this section, we first introduce some basic concepts in legal case retrieval. Then we delve into the three core parts of our KELLER, including legal knowledge-guided case reformulation, relevance modeling, and dual-level contrastive learning.

\subsection{Preliminaries}

In legal case retrieval,
both queries and candidate documents are real structured legal cases that can extend to thousands of tokens in length. Figure~\ref{fig:Fig_CaseStructure} shows an illustration of the typical case structure. 
Specifically, a case usually contains several sections, including \textit{procedure}, \textit{fact}, \textit{reasoning}, \textit{decision}, and \textit{tail}. 
Notably, the candidate documents are completed legal cases that have been through the adjudication process and therefore contain all sections. In contrast, the query cases are not yet adjudicated, so they usually only include the \textit{procedure} and \textit{fact} sections.

Formally, given a query case $q$ and a set of document cases $D$, the objective of legal case retrieval is to calculate a relevance score $s$ between the query case and each document case in $D$, and then rank the document cases accordingly.

% In most practical systems, the formats of query cases and candidate cases are not identical. Typically, candidate cases are legally adjudicated cases, encompassing comprehensive content such as facts, evidence, reasoning, and judicial decisions. In contrast, queries are the cases submitted by the procuratorate to the court, usually containing only the alleged crimes, corresponding facts, and evidence. Our objective is to retrieve cases that are most similar to the given scenario, thereby providing professional users with essential support for downstream tasks.

\subsection{Knowledge-Guided Case Reformulation}
When assessing the relevance between two legal cases, the key facts of their crimes are the most crucial things for consideration.
Therefore, given the complexity of the original legal cases which makes direct learning challenging, we try to first refine the legal cases into shorter but more essential ``crime-fact'' snippets. For example, we can get such a snippet from the case shown in Figure~\ref{fig:Fig_CaseStructure}, whose crime is \textit{``the crime of arson''} and the fact is \textit{``Yan took advantage of Mu's absence and set fire ...''}.

However, the description of a crime and its corresponding facts are often scattered throughout the lengthy case, and a single case may contain multiple crimes and facts, significantly complicating the extraction process. 
% Moreover, LLMs are typically developed for general scenarios and lack the specialization to identify legal facts that should be summarized.
To tackle this problem, we propose a two-step prompting method leveraging professional legal knowledge to guide LLM to achieve accurate extraction. \\

\noindent \textbf{Crime and law article extraction.}
First, we prompt LLM to extract all crimes and all law articles from the case. This step is relatively straightforward for LLM, as each crime and law article is a distinct, identifiable element within the text. For example, the extracted crime and law article for the case shown in Figure~\ref{fig:Fig_CaseStructure} are \textit{``the crime of arson''} and \textit{``Article 114 and Paragraph 1 of Article 67 of the Criminal Law of the People's Republic of China''}, respectively.
Our extraction prompt is shown in Appendix~\ref{sec:appendix_extraction_prompt}. \\

\noindent \textbf{Post-Processing.}
The extracted law articles may just be the titles. We then expand these titles into full articles by gathering their detailed provision content from the Web based on the titles.
Then, we establish a mapping between each crime and its relevant law articles by referring to a database built by our legal experts.
Note that the correlation between specific crimes and their corresponding legal articles is objective, as it is clearly defined by law.
After post-processing, we can obtain all the ``crime-articles'' pairs for a legal case. \\

\noindent \textbf{Fact summarization.}
Next, we leverage the extracted crimes and their relevant law articles to guide LLM in summarizing the specific facts of each crime from the original legal case. The law articles, serving as high-level abstractions of the actual criminal events, can considerably simplify the task of identifying the corresponding specific facts. 
The prompt for fact summarization is shown in Appendix~\ref{sec:appendix_summarization_prompt}. \\

Through our legal knowledge-guided reformulation, we can accurately distill a series of crimes and their corresponding specific facts from the originally lengthy legal cases.
Finally, we form a \textit{sub-fact} snippet, with the crime as the title and its facts as the main body. These refined sub-facts are not only the most crucial information for relevance judgment in legal case retrieval but are also concise enough to meet the text length limitations of popular pre-trained retrieval models. Please note that, since the required legal knowledge is present in criminal case documents from mainstream countries (e.g., China and the United States), our approach is actually internationally applicable. Our materials in Appendix~\ref{Appendix: case format} further prove this.

\subsection{Relevance Modeling}
We directly model the relevance of legal cases using the refined sub-facts, rather than relying on the full text of the original legal cases.
Specifically, given a query case $q=\{q_1, ..., q_m\}$ and a candidate case $d=\{d_1, ..., d_n\}$, where $q_i$ represents the $i$-th  sub-fact of $q$ and $d_j$ represents the $j$-th sub-fact of $d$. We utilize a pre-trained text encoder to encode them:
\begin{equation}
    \begin{aligned}
        E_{q_i} & = \text{Pool}_{\text{[CLS]}}\left(\text{Encoder}(q_i)\right),\\
        E_{d_j} & = \text{Pool}_{\text{[CLS]}}\left(\text{Encoder}(d_j)\right),
    \end{aligned}
\end{equation}
where $\text{Pool}_{\text{[CLS]}}$ means extracting the embedding output at the [CLS] token position. 
Then, we compute the similarity matrix $\mathbf{M}_{m \times n}$ using the L2-norm dot product. Each element $M_{i, j}$ of $M$ is the similarity calculated between the normalized embeddings of the $i$-th sub-fact in the reformulated query case and $j$-th sub-fact in the reformulated document case:
\begin{equation}
    \begin{aligned}
    M_{i, j} = \text{Sim}(E_{q_i},E_{d_j}) = \text{Norm}(E_{q_i}) \cdot \text{Norm}(E_{d_j}^T).
    \end{aligned}
\end{equation}

Finally, we aggregate this similarity matrix to derive the matching score.
There are various sophisticated choices for aggregation, such as using attention or kernel pooling~\cite{KNRM}. In this paper, we opt to employ the \textit{MaxSim} and \textit{Sum} operators~\cite{khattab2020colbert}:
\begin{equation}
    \begin{aligned}
    s_{q, d} = \sum_{i=1}^{m} \text{Max}_{j =1}^{n} M_{i, j},
    \end{aligned}
\end{equation}
where $s_{q, d}$ is the final predicted relevance score. 
We choose these two operators because of their advantages in effectiveness, efficiency, and interpretability over the other aggregation approaches for our scenario:

\textbf{(1) Effectiveness}: Typically, each query's sub-fact $q_i$ matches one document sub-fact $d_j$ at most in practice, which is well-suited for \textit{MaxSim} of applying the Max operation across all document's sub-facts for a given query's sub-fact.
For instance, considering a query sub-fact about \textit{``drug trafficking''}, and the document sub-facts about \textit{``drug trafficking''} and \textit{``the discovery of privately stored guns and ammunition''},  only the \textit{``drug trafficking''} sub-fact of the document is relevant for providing matching evidence. In contrast, using soft aggregation methods (e.g., kernel pooling~\cite{KNRM}) may introduce additional noise in this scenario.

\textbf{(2) Efficiency}: \textit{Maxsim} and \textit{Sum} operations on tensors are quite efficient for both re-ranking and large-scale top-\textit{k} retrieval supported by multi-vector-based Approximate Nearest Neighbor algorithms~\cite{khattab2020colbert}. This high efficiency is important for meeting the low-latency requirements of the practical use.
% of legal case retrievers.
% \textcolor{blue}{[This is important for the real practice of legal case retrieval......]}

\textbf{(3) Interpretability}: \textit{MaxSim} provides clear interpretability by revealing the quantitative contribution of each query and document sub-fact towards the final relevance score, which can aid in understanding the ranking strategies and justifying the retrieval results. We further illustrate this advantage by studying a real case in Section~\ref{sec: case study}.

% \begin{itemize}[leftmargin=*]
%   \item \textbf{(1) Efficiency:} The MaxSim operator involves only the Max and Sum operations on tensors, resulting in rapid processing speeds. Furthermore, it facilitates direct end-to-end top-K retrieval via pruning, rather than merely reranking.
%   \item \textbf{(2) Interpretability:} Leveraging the MaxSim operator, we can readily ascertain the extent to which each $q_i$ is satisfied, and identify which $d_j$ contributes matching evidence to $q_i$.
%   \item \textbf{(3) Effectiveness:} During the case reformulation phase, the case is decomposed into multiple, relatively independent facts. In most cases, the information need in each $q_i$ is at most satisfied by one $d_j$. Consider an example where the query pertains to a drug trafficking incident, while the candidate cases involve both drug trafficking and the discovery of privately stored guns and ammunition. In such scenarios, only the drug trafficking aspect of the case is relevant for providing matching evidence. Employing soft aggregation methods like the attention layer might introduce additional noise and reduce performance, which can be observed in the experimental results.
% \end{itemize}

% The capability of the text encoder to estimate the semantic relevance between criminal facts is crucial for the proposed model architecture. We design two contrastive learning objectives at different levels of granularity, aimed at enhancing our model's ability from multiple perspectives.

\subsection{Dual-Level Contrastive Learning}
We incorporate matching signals from both the coarse-grained case level and the fine-grained sub-fact level to comprehensively enhance the model performance in legal case matching. \\

\noindent \textbf{Case-level contrastive learning.} 
At the case level, we consider directly optimizing toward the final matching score between the query case and the document cases. Specifically, we employ the classical ranking loss function to promote the relevance score between the query and the positive document while reducing it for negative documents:
\begin{equation}
    \begin{aligned}
    \mathcal{L}_{\text{R}} = -\log \frac{\exp(s_{q,d^+}/\tau)}{\exp(s_{q,d^{+}}/\tau) + \sum_{d^-} \exp(s_{q,d^{-}}/\tau)},
    \end{aligned}
\end{equation}
where $d^+$ is the positive document of the query $q$ and each $d^{-}$ is from the in-batch negatives. $\tau$ is a temperature parameter. \\

\noindent \textbf{Sub-fact-level contrastive learning.} 
At the sub-fact level, we incorporate intermediate relevance signals among sub-facts to fine-grainedly enhance the model's effectiveness in understanding sub-facts' content and their matching relationships. 
However, only the case-level relevance labels are available in the dataset. Naively considering all the sub-fact pairs between the query and the positive documents as positives and all the sub-fact pairs between the query and the negative documents as negatives will introduce substantial false positive and negative noise.
To mitigate this issue, we propose a heuristic strategy to obtain high-quality relevance labels for the query's sub-facts $\{q_1, ..., q_m\}$. 
The core idea of this strategy is to combine the case-level relevance and the charges of each sub-fact to accurately identify true positive and negative samples.
We introduce the details of this strategy in Appendix~\ref{sec:appendix_relevance_label_strategy} due to the space limitation.

After getting the sub-fact level relevance labels, we also adopt the ranking loss function for sub-fact level contrastive learning:
\begin{equation}
    \begin{aligned}
    \mathcal{L}_{\text{S}} = -\log \frac{\exp(s_{M_{i,j^{+}}}/\tau)}{\exp(s_{M_{i,j^{+}}}/\tau) + \sum_{J^{-}} \exp(s_{M_{i,j^{-}}}/\tau)},
    \end{aligned}
\end{equation}
where $M_{i, j^{+}}$ are the similarity score between $q_i$ and its positive document. $M_{i, j^{-}}$ are the similarity score between $q_i$ and its negative document sub-fact.
$J^{-}$ is the collection of all negative document sub-facts for $q_i$.
The final learning objective is the combination of $\mathcal{L}_{\text{R}}$ and $\mathcal{L}_{\text{S}}$:
\begin{equation}
    \begin{aligned}
        \mathcal{L} = \mathcal{L}_{\text{R}} + \alpha \mathcal{L}_{\text{S}},
    \end{aligned}
\end{equation}
where $\alpha$ is a hyper-parameter to adjust the weights of two losses.

% We define a Mask operation: similarity scores of positive/negative samples that do not satisfy the above two rule-based strategies will be set to $-\inf$; otherwise, they will retain their original values. \deng{The equation needs to be modified.} Then, the loss function of the sub-fact-level contrastive learning can be defined as follows:
% \begin{equation}
%     \begin{aligned}
%     \mathcal{L}_{crime} &= -\sum_{i=0}^{|E_q|} \log P_i \\
%     P_i &= \begin{cases}
%     \frac{\hat{S}_{q_i, d^+_k}}{\hat{S}_{q_i, d^+_k} + \sum_{j \neq k}\hat{S}_{q_i, d^+_j} + \sum_{d^-} \sum_{j}\hat{S}_{q_i, d^-_j}}, & \text{if } \hat{S}_{q_i, d^+_k} \neq 0 \\
%     1, & \text{otherwise}
%   \end{cases}, \\
%   \hat{S}_{q_i, d_j} &= \exp(S_{q_i, d_j}/\tau)
%     \end{aligned}
% \end{equation}

\section{Experiments}
\subsection{Experimental Setup}
\noindent \textbf{Dataset and evaluation metrics.}
We conduct extensive experiments on two widely-used datasets: LeCaRD~\cite{ma2021lecard} and LeCaRDv2~\cite{li2023lecardv2}, whose statistics are listed in Appendix~\ref{sec:appendix-dataset}.
Considering the limited number of queries in LeCaRD, we directly evaluate all the queries of LeCaRD using the best model trained on LeCaRDv2, thereby avoiding the need for dataset split.
% LeCaRD comprises 107 queries and 10,700 candidate cases. LeCaRDv2, a more extensive collection, includes 800 queries and 55,192 candidate cases. 
% Both datasets employ a four-level relevance annotation principle, ranging from 0 to 3, with increasing levels of relevance. 
Following the previous studies~\cite{DBLP:conf/sigir/LiACDW0CT23, li2023lecardv2}, we regard label=3 in LeCaRD and label$\geq$2 in LeCaRDv2 as positive. For the query whose candidate documents are all annotated as positive, we supplement the candidate pool by sampling 10 document cases from the top 100-150 BM25 results. 
% We follow the official split of the query set for LeCaRDv2, where the training set comprises 640 queries, and the test set includes 160 queries. 
% Considering the limited number of queries in LeCaRD, we directly evaluate all the queries of LeCaRD using the best model trained on LeCaRDv2, thereby avoiding the need for dataset split.
To exclude the effect of unlabeled potential positives in the corpus, we rank the candidate pools and adopt MAP, P@k (k=3), and NDCG@k (k=3, 5, 10) as our evaluation metrics.  \\
% Given the presence of detailed four-level relevance annotations in both datasets, we apply the exponential decay function to assign relevance weights in the NDCG@k metric. Specifically, the weight assigned to each candidate case is defined as $2^{n-1}$, where $n$ represents the relevance level annotated in the dataset. Notably, the corresponding weight is set to 0 when n=0.

\begin{table*}[]
\renewcommand\arraystretch{1.05}
% \small
%\small %dou: \small is the last choice
\centering
\caption{Main results of the fine-tuned setting on LeCaRD and LeCaRDv2. ``$\dagger$'' indicates our approach outperforms all baselines significantly with paired t-test at p $<$ 0.05 level. The best results are in bold.}\label{table:main results}
\scalebox{0.75}{\begin{tabular}{l|ccccc|ccccc}
\toprule
\multirow{2}{*}{Model} & \multicolumn{5}{c|}{LeCaRD} & \multicolumn{5}{c}{LeCaRDv2} \\
 & MAP & P@3 & NDCG@3 & NDCG@5 & NDCG@10 & MAP & P@3 & NDCG@3 & NDCG@5 & NDCG@10 \\
 \hline
 \multicolumn{11}{c}{\textit{Traditional ranking baselines}} \\
BM25  & 47.30 & 40.00 & 64.45 & 65.59 & 69.15 & 55.20 & 48.75 & 72.11 & 72.51 & 79.85\\
TF-IDF & 42.59 & 36.19 & 58.14 & 59.98 & 63.37 & 55.19 & 47.92 & 71.38 & 72.70 & 75.04 \\
\hline
 \multicolumn{11}{c}{\textit{PLM-based neural ranking baselines}} \\
BERT  & 53.83 & 50.79 & 73.19 & 73.43 & 75.54 & 60.66 & 53.12 &  77.78 & 78.73 & 80.85 \\
RoBERTa & 55.79 & 53.33 & 74.40 & 74.33 & 76.70 & 59.75 &  53.12 & 78.15 & 78.97 & 80.70 \\
BGE & 54.98 & 53.33 & 74.29 & 74.09 & 75.65 & 60.64 & 51.87 & 76.99 & 78.43 & 80.90 \\
SAILER & 57.98 & 56.51 & 77.55 & 77.04 & 79.41 & 60.62 & 54.58 & 78.67 & 78.99 & 81.41 \\
\hline
 \multicolumn{11}{c}{\textit{Neural ranking baselines designed for long text}} \\
BERT-PLI & 48.16 & 43.80 & 65.74 & 68.14 & 71.32 & 55.34 & 46.67 & 71.62 & 73.68 & 76.63 \\
Lawformer & 54.58 & 50.79 & 73.19 & 73.43 & 75.54 & 60.17 & 54.17 & 78.23 & 78.99 & 81.40 \\
\hline
\multicolumn{11}{c}{\textit{Case reformulation with LLMs}} \\
PromptCase & 59.71 & 55.92 & 78.75 & 78.44 & 80.71 & 62.25 & 54.19 & 78.51 & 79.07 & 81.26\\
KELLER & \textbf{66.84}$^\dagger$ & \textbf{57.14} & \textbf{81.24}$^\dagger$ & \textbf{82.42}$^\dagger$ & \textbf{84.67}$^\dagger$ & \textbf{68.29}$^\dagger$ & \textbf{63.13}$^\dagger$ & \textbf{84.97}$^\dagger$ & \textbf{85.63}$^\dagger$ & \textbf{87.61}$^\dagger$ \\
\bottomrule
\end{tabular}}
\end{table*}

\noindent \textbf{Baselines.}
We compare KELLER against the following baselines across three categories.
The first is \textit{traditional probabilistic models}, including {TF-IDF} and {BM25}.
The second is \textit{ranking methods based on pre-trained language models}, including {BERT}~\cite{bert}, {RoBERTa}~\cite{DBLP:journals/corr/abs-1907-11692},  {BGE}~\cite{DBLP:journals/corr/abs-2309-07597} and {SAILER}~\cite{DBLP:conf/sigir/LiACDW0CT23}.
The third is \textit{ranking methods designed for handling long (legal) text}, including {BERT-PLI}~\cite{shao2020bert}, {Lawformer}~\cite{xiao2021lawformer}, and {PromptCase}~\cite{DBLP:conf/adc/TangQL23}. \\

\noindent \textbf{Implementations.}
We introduce the selected language models, hyperparameter settings and other details in Appendix~\ref{sec:appendix-implementation}.

\subsection{Main Results}
The main results are as shown in Table~\ref{table:main results} and we have the following observations:

\textbf{(1) KELLER outperforms all baseline methods across all metrics on both datasets.} Compared with previous methods tailored for the long-text problem, KELLER employs knowledge-guided case reformulation to address the challenge of long-text comprehension. This demonstrates the effectiveness of separating comprehension and matching tasks in the domain of legal case retrieval.

\textbf{(2) After fine-tuning on legal case retrieval datasets, the performance gap between general-purpose and retrieval-oriented PLMs becomes less distinct.} This observation may stem from two reasons. First, the scarcity of training data in the legal case retrieval task can induce overfitting to annotation signals, which hampers the model's generalization capabilities. Second, Naive truncation of lengthy texts can make the model's inputs lose sufficient matching signals, leading to inconsistencies between relevance annotations and matching evidence.

\textbf{(3) We observe that these long-text-oriented baseline methods do not show significant advantages.}  Despite BERT-PLI and Lawformer processing more text than other methods, their input capacity was still insufficient for the average length of legal cases. Handling both long-text processing and complex semantic understanding within one retriever presents a significant challenge. To address this issue, our approach offloads a portion of the long-text comprehension task via knowledge-guided case reformulation and improves the ranking performance.

% \textbf{(4)  Pre-trained methods generally outperform traditional lexical methods after fine-tuning.} This indicates that, despite the negative impact of long and complex texts in legal case retrieval, pre-trained methods can still effectively capture textual semantics after task adaptation.

\begin{table*}[!t]
\renewcommand\arraystretch{1.1}
\centering
\caption{Zero-shot performance on LeCaRD and LeCaRDv2. ``$\dagger$'' indicates our approach outperforms all baselines significantly with paired t-test at p $<$ 0.05 level. The best results are in bold.}\label{table:zeroshot}
\scalebox{0.75}{\begin{tabular}{l|ccccc|ccccc}
\toprule
\multirow{2}{*}{Model} & \multicolumn{5}{c|}{LeCaRD} & \multicolumn{5}{c}{LeCaRDv2} \\
 & MAP & P@3 & NDCG@3 & NDCG@5 & NDCG@10 & MAP & P@3 & NDCG@3 & NDCG@5 & NDCG@10 \\
\hline
 \multicolumn{11}{c}{\textit{General PLM-based baselines}} \\
BERT & 42.92 & 37.78 & 60.11 & 61.37 & 64.10 & 56.46 & 
52.08 & 75.82 & 77.05 & 79.39 \\
RoBERTa & 51.50 & 47.62 & 69.21 & 71.07 & 73.60 & 57.89 & 52.08 & 75.48 & 76.33 & 78.38 \\
Lawformer & 42.80 & 38.41 & 59.46 & 61.61 & 64.13 & 55.05 & 49.58 & 74.42 & 74.31 & 76.96 \\
\hline
 \multicolumn{11}{c}{\textit{Retrieval-oriented pre-training baselines}} \\
BGE & 51.81 & 47.62 & 68.57 & 69.91 & 72.61 & 57.21 & 
50.42 & 73.59 & 75.36 & 77.80 \\
SAILER & 60.62 & 56.19 & 79.93 & 78.99 & 81.41 & 62.80 & 
55.00 & 79.38 & 81.17 & 83.83 \\
\midrule
KELLER & \textbf{64.17}$^\dagger$ & \textbf{57.78} & \textbf{80.47} & \textbf{81.43}$^\dagger$ & \textbf{84.36}$^\dagger$ & \textbf{65.87}$^\dagger$ & \textbf{61.67}$^\dagger$ & \textbf{83.33}$^\dagger$ & \textbf{83.75}$^\dagger$ & \textbf{86.06}$^\dagger$ \\
\bottomrule
\end{tabular}}
\end{table*}

\subsection{Zero-shot Evaluation}
% In recent years, pre-trained models have exhibited superior zero-shot retrieval performance. 
Considering the inherent data scarcity problem in legal case retrieval, 
we evaluate the zero-shot performance (i.e., without fine-tuning on the training set of LeCaRDv2) of models on LeCaRDv2. 

Results are shown in Table~\ref{table:zeroshot} and we find that {KELLER consistently outperforms baselines in both zero-shot and fine-tuning settings.} 
Upon comparing the performance of each method under zero-shot and fine-tuned settings, we observe that most methods benefit from fine-tuning except SAILER. Intuitively, models trained in a general domain or task could be enhanced through fine-tuning. In specific domains, continued fine-tuning of models generally does not lead to a significant decrease in performance. We posit that the unexpected outcomes in the SAILER model primarily arise from overfitting the limited data used for fine-tuning, which impairs the generalization capabilities established in the pre-training phase.

% Despite sharing the same text encoder as SAILER, KELLER exhibits enhancements after fine-tuning. 

% (2) \textbf{Retrieval-oriented pre-trained models (BGE and SAILER) outperform BERT}, which indicates that pre-training focused on retrieval tasks is effective in the domain of legal case retrieval. Surprisingly, RoBERTa achieves comparable performance to BGE, which is consistent with the previous work~\cite{DBLP:conf/sigir/LiACDW0CT23}. On the other hand, SAILER demonstrates a clear improvement over BGE. This enhancement can likely be attributed to the fact that SAILER undergoes pre-training on legal documents, thereby eliminating the need for further domain adaptation.

\subsection{Ablation Study} \label{sec:Ablation}
We design the following six ablations:
(1) \textit{KGCR$\rightarrow$NS}: We replace our Knowledge-Guided Case Reformulation (KGCR) with a Naive Summarization (NS), which produces case summaries without hierarchical structure. We subsequently optimize the dual encoders with this text as the input. 
(2)\textit{MS $\rightarrow$ Mean}: We replace \textit{MaxSim} and \textit{Sum} (MS) with  \textit{Mean} to capture the average relevance of each sub-fact in the candidate cases to the query.
(3) \textit{MS $\rightarrow$ NC}: We Naively Concatenate (NC) all the reformulated sub-facts into a text sequence and subsequently optimize the dual-encoders.
(4) \textit{MS $\rightarrow$ KP}: We employ kernel pooling~\cite{KNRM} on the score matrix to capture relevance signals.
(5) \textit{w/o sfCL}: Training without the sub-fact-level contrastive learning.
(6) \textit{w/o SfCL}: Training without the case-level contrastive learning.

Results are shown in Table~\ref{table:ablation} and we can observe:

(1) Every ablation strategy results in a decline in the model's performance, demonstrating the effectiveness of each module within KELLER. This outcome indicates that KELLER's architecture is both comprehensive and synergistic, with each module contributing to the model's overall performance.

(2) The replacement of the KGCR module exhibits the most significant impact on performance. This highlights the pivotal role of the KGCR module in KELLER. The KGCR module decomposes cases into structured sub-facts, which are crucial for the model's learning process.

(3) Among different aggregation strategies, \textit{MS $\rightarrow$ Mean} demonstrates the least performance degradation. This is primarily because the dataset mainly consists of simple cases with single charges, where  \textit{Mean} and  \textit{MS}  become essentially equivalent. Conversely,  \textit{MS $\rightarrow$ NC}  exhibits the most notable performance decline. This is mainly because the model no longer maintains a cross-matching architecture after the concatenation operation. Merging multiple facts into a single representation negatively impacts representation learning.

% (4) Surprisingly, in the absence of the case-level relevance signals provided by the dataset (i.e., \textit{w/o CaCL}), KELLER achieves comparable performance solely relying on sub-fact-level self-supervised signals. This demonstrates the importance of modeling intermediate relevance signals in complex legal case scenarios.

\begin{table}[!t]
\centering
\caption{Results of ablation study on LeCaRDv2.}\label{table:ablation}
\renewcommand\arraystretch{1.2}
\resizebox{0.94\linewidth}{!}{%
    \begin{tabular}{l|ccccc}
    \toprule
    Strategy & MAP & P@3 & NDCG@3 & NDCG@5 & NDCG@10\\
    \hline
     \multicolumn{6}{c}{\textit{{Effect of knowledge-guided case reformulation}}} \\
    \textit{KGCR}$\rightarrow$\textit{NS} &61.91 & 55.13 & 79.50 & 79.11 & 81.47\\
    \hline
     \multicolumn{6}{c}{{\textit{Effect of different aggregation strategy}}} \\
  
    \textit{MS}$\rightarrow$\textit{Mean} & 67.15 & 61.81 & 81.58 & 84.42 & 86.74 \\
    \textit{MS}$\rightarrow$\textit{NC} & 63.35 & 57.92 & 80.37 & 81.99 & 84.04\\
    \textit{MS}$\rightarrow$\textit{KP} & 65.47 & 60.06 & 79.87 & 83.61 & 85.39 \\
    \hline
     \multicolumn{6}{c}{{\textit{Effect of contrastive learning}}} \\
    \textit{w/o} \textit{SfCL} & 67.39 & 61.93 & 81.24 & 84.73 & 86.91 \\
    \textit{w/o} \textit{CaCL} & 67.18 & 61.67 & 82.76 & 84.45 & 86.51\\
    \midrule
    KELLER & \textbf{68.29} & \textbf{63.13} & \textbf{84.97} & \textbf{85.63} & \textbf{87.61} \\
    \bottomrule
\end{tabular}
} 
\end{table}

\subsection{Evaluations on Different Query Types}

\begin{figure}[!t]
	\centering
	\includegraphics[width=0.95\linewidth]{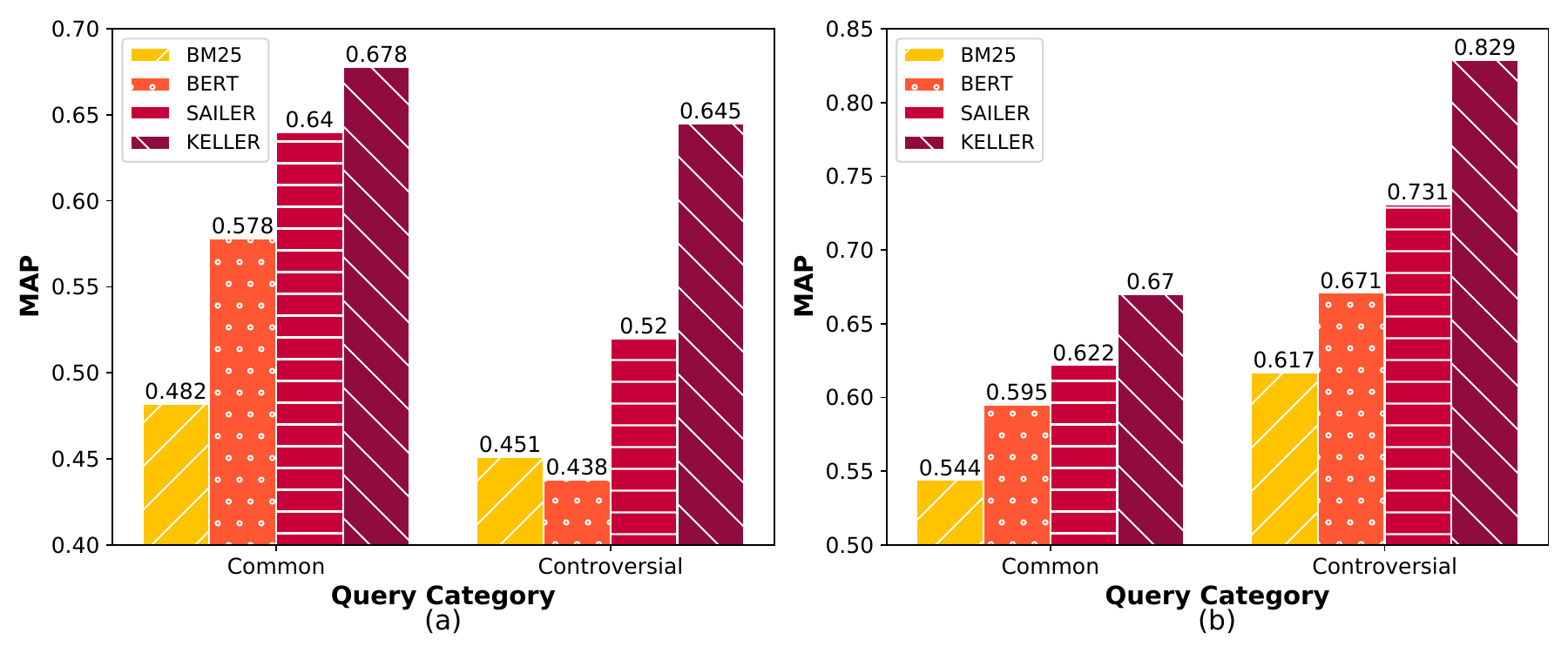}
	\caption{Evaluation on different query types. We evaluate four models on (a) LeCaRD and (b) LeCaRDv2.}
	\label{fig:Fig_query_types}
\end{figure}

\begin{figure*}[!t]
	\centering
	\includegraphics[width=0.96\linewidth]{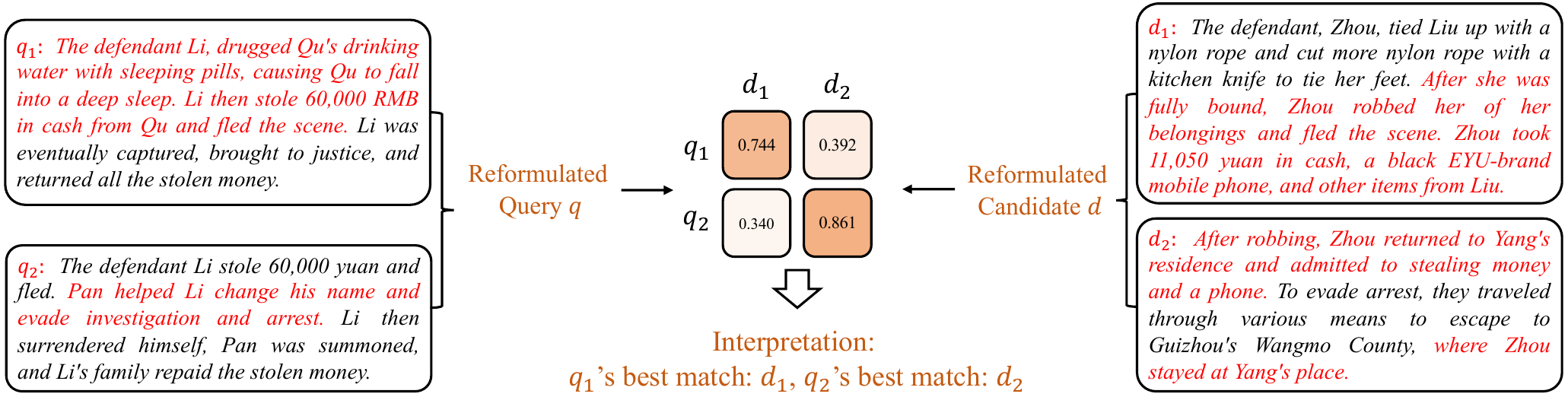}
	\caption{An example of the interpretability of KELLER. We can observe that each sub-fact of the query finds a correct match in the candidate document (in red).}
	\label{fig:Fig_CaseStudy_Interpretability}
\end{figure*}

\begin{figure}[!t]
	\centering
	\includegraphics[width=1\linewidth]{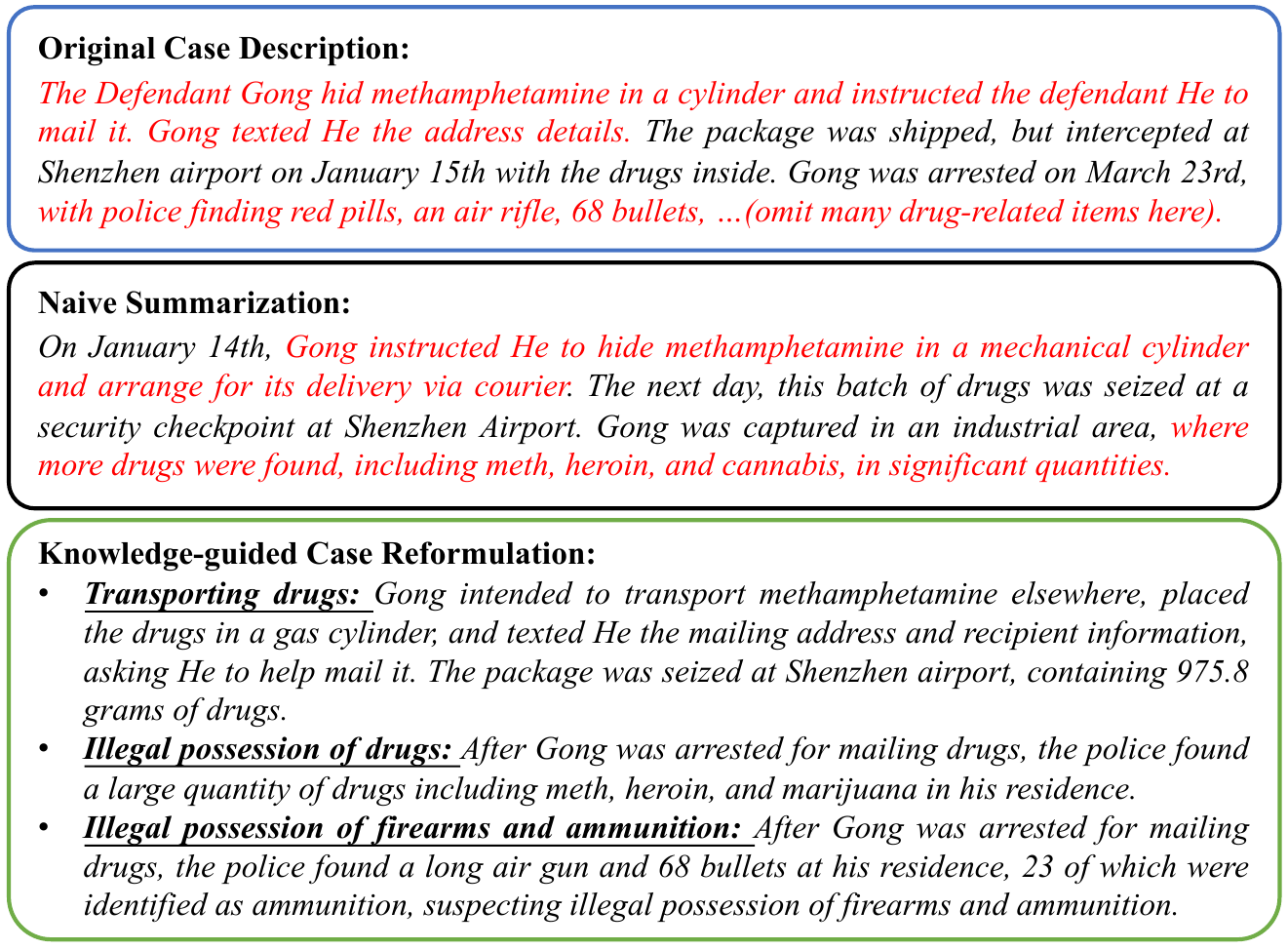}
	\caption{Comparison of the original text, naive summarization, and our proposed knowledge-guided case reformulation. The original text is manually abbreviated due to its length. Important sentences are marked in red.}
	\label{fig:Fig_CaseStudy_CaseReformulation}
\end{figure}

We investigate the two query types presented in both LeCaRD and LeCaRDv2: \textit{common} and \textit{controversial}. Common queries are similar to initial trials, and controversial queries to retrials, which are typically more complex and require additional expert review. We evaluated multiple models on these query types. Notably, SAILER's performance declined after fine-tuning, so we included its zero-shot results for comparison, alongside the fine-tuned outcomes of other models. Results as shown in Figure~\ref{fig:Fig_query_types} and we find:

(1) KELLER outperformed other models on both query types, showing more substantial gains in controversial queries with improvements of 24.04\% and 13.41\% in the LeCaRD and LeCaRDv2 datasets, respectively. This enhanced performance is credited to KELLER’s novel case reformulation, which simplifies complex scenarios into sub-facts, aiding in better comprehension and matching.

(2) In the LeCaRD dataset, lexical-based models showed consistent performance across different queries, unlike representation-based models which varied significantly. For example, BERT outperformed BM25 on common queries but was less effective on controversial ones, a difference attributed to the models’ limited ability to handle multifaceted cases. KELLER’s cross-matching architecture successfully addresses this limitation.

\subsection{Case Studies}
\label{sec: case study}
\noindent \textbf{Case reformulation.}
% The objective of case reformulation is to present clear texts with matching evidence for the retriever. Previous experimental results in Section~\ref{sec:Ablation} have demonstrated that our knowledge-guided case reformulation outperforms naive summarization in effectiveness. 
We provide an illustrative comparison between the original case description, naive summarization, and our knowledge-guided case reformulation in Figure~\ref{fig:Fig_CaseStudy_CaseReformulation}. The case centers on complex issues of drug transport and firearm possession. Most details focus on drug transportation, with brief mentions of firearms found at the defendant's residence towards the end. Given the 512-token limit of most retrievers, crucial information about the firearms is often inaccessible. While naive summarization captures the main points, it overlooks specifics about the firearms in the context of drug offenses. In contrast, our KGCR method segments the case into three topics—drug transportation, illegal drug possession, and illegal firearms possession—thus detailing each criminal aspect comprehensively. \\

\noindent \textbf{Interpretability.}
% Compared to the Dual-Encoder and Cross-Encoder architectures~\cite{PLM_IR_Survey}, KELLER leverages a reformulated case structure to achieve intuitive interpretability.
In KELLER, each sub-fact in a query represents a specific intent of the query, with the highest match score from a candidate case indicating how well this intent is met. KELLER allows users to see which sub-fact in a candidate case matches their intent. For example, in a case involving robbery and harboring crimes shown in Figure~\ref{fig:Fig_CaseStudy_Interpretability}, KELLER accurately matches sub-facts in the query to those in the candidate case, demonstrating the alignment of KELLER’s scoring with the underlying legal facts of the case. The matching is shown in a matrix, where the positions $(q_1, d_1)$ and $(q_2, d_2)$ highlight the defendant's actions in the query and the candidate case, respectively, establishing a direct correlation between the computed scores and the case ranking.

\section{Conclusion}
In this paper, we introduce KELLER, a ranking model that effectively retrieves legal cases with high interpretability. KELLER structures legal documents into hierarchical texts using LLMs and determines relevance through a cross-matching module. Our tests on two expert-annotated datasets validate its effectiveness. In the future, we will enhance KELLER by incorporating additional specialized knowledge and generative models to refine performance and produce language explanations.

% \clearpage
\section{Limitations}
\noindent \textbf{External Knowledge base Construction.}
Our method requires constructing a legal knowledge base to assist in case reformulation, which introduces an extra step compared to the out-of-the-box dense retrievers. This issue is common in most domain-specific knowledge-enhanced methods.

\noindent \textbf{Computing Efficiency.}
Our approach needs to call large language models when processing the query case, which may bring additional computational costs. In our experiments, we have employed techniques such as vLLM to achieve high-speed inference. Furthermore, we believe that with ongoing advancements in techniques in both hardware and algorithms, the computational of utilizing LLMs for processing individual query cases online will be acceptable. For example, Llama3-8B can achieve a speed exceeding 800 tokens per second on the Groq platform, while recent inference services provided by Qwen and DeepSeek require less than \$0.0001 per 1,000 tokens.

% Our approach employs large language models to reformulate legal cases, which may be limited by two factors: (1) The quality of case reformulation depends on the capabilities of the large language models used. Our experiments in this paper employ Qwen-72B-Chat, a leading open-source LLM in Chinese, which can follow instructions and accurately extract targeted content. However, models with smaller sizes (e.g., 7B) struggle to perform this task effectively. (2) The efficiency of online inference is highly influenced by the deployment conditions of the LLM. Enhanced inference latency can be achieved through hardware adaptations or advanced inference architectures such as vLLM.

\section{Ethical Discussion}
% Legal case retrieval is an emerging and increasingly prominent field within the realm of intelligent judicial systems. However, 
The application of artificial intelligence in the legal domain is sensitive, requiring careful examination and clarification of the associated ethical implications. The two datasets utilized in our experimental analysis have undergone anonymization processes, particularly with regard to personally identifiable information such as names. 

Although KELLER demonstrates superior performance on two human-annotated datasets, its recommendations for similar cases may sometimes be imprecise when dealing with intricate real-world queries. Additionally, the case databases in existing systems may not consistently include cases that fully satisfy user requirements. 
% We hope that this technology can assist professional users in locating similar cases. Nonetheless, 
The choice to reference the retrieved cases should remain at the discretion of the experts.

% \clearpage

\appendix

\section{More Details for Experimental Setup}
\subsection{Datasets}
\label{sec:appendix-dataset}
\begin{table}[!t]    
    \caption{Basic statistics of the datasets.}
    \label{table:dataset}
    \scalebox{0.82}{\begin{tabular}{p{0.3\textwidth}p{0.1\textwidth}p{0.1\textwidth}}
        \toprule
        Dataset & LeCaRD & LeCaRDv2 \\
        \midrule
        \# Train queries & - & 640\\
        \# Test queries & 107 & 160\\
        \# Documents & 9,195 & 55,192\\
        Average query length & 445 & 4,499\\
        Average doc length & 7,446 & 4,768\\
        Average golden docs / query & 10.39 & 13.65\\  
        \bottomrule
    \end{tabular}}
\end{table}

The statistics of both datasets are listed in Table~\ref{table:dataset}. LeCaRD comprises 107 queries and 10,700 candidate cases. LeCaRDv2, a more extensive collection, includes 800 queries and 55,192 candidate cases.

\subsection{Implementation Details}
\label{sec:appendix-implementation}
For baseline models, we employ the default parameter settings of Okapi-BM25 in the implementation of BM25. For ranking methods based on PLMs, a uniform learning rate of 1e-5 and a batch size of 128 are consistently applied. In BERT-PLI, the numbers of queries and candidate case segments are set to 3 and 4, respectively, with a maximum segment length of 256. For Lawformer, the maximum text input length is set to 3,072, optimized using a learning rate of 1e-5 and a batch size of 64.

%~\footnote{\url{https://chat.openai.com/}}
In KELLER, we employ the Qwen-72B-Chat~\cite{bai2023qwen}, which is currently one of the best open-source Chinese LLMs, to perform case reformulation. 
We do not choose OpenAI API due to concerns about reproducibility and high cost. 
All prompts, except for the case description, are input as system prompts. In the ranking model, the maximum number of crimes per case is capped at 4, which meets the needs of most cases. We adopt the pre-trained retriever SAILER as the text encoder. The $\tau$ in the contrastive learning is 0.01, and the $\alpha$ in the final loss function is 0.9. We conduct model training with a learning rate of 1e-5 and a batch size of 128. All experiments are conducted on four Nvidia Tesla A100-40G GPUs. 
% All the source code and data will be shared at \url{https://github.com/hide-for-blind-review} if the paper gets accepted.

\section{Prompts}
\label{sec:appendix_extraction_prompt}
\subsection{Extraction Prompt}

\begin{center}
    \begin{minipage}{.45\textwidth}
    \textbf{Extraction Prompt:} \textit{You are now a legal expert, and your task is to find all the crimes and law articles in the procuratorate's charges (or court judgments) from the provided case. The output format is one line each for crimes and law articles, two lines in total. Multiple crimes (law articles) are separated by semicolons.}
    \end{minipage}
\end{center}

\subsection{Summarization Prompt}
\label{sec:appendix_summarization_prompt}
\begin{center}
    \begin{minipage}{.45\textwidth}
    \textbf{Summarization Prompt:} \textit{You are now a legal expert, and you are good at analyzing lengthy legal case texts containing multiple circumstances of crime. Your task is to concisely summarize the causes, procedures, and outcomes associated with a specified crime, ensuring each part does not exceed 100 words.}

    [\textit{Crime}]: \textit{the specific crime name}

    [\textit{Law Articles}]: \textit{the specific provisions of law articles}
    \end{minipage}
\end{center}

\section{Strategy to Obtain Sub-Fact-Level Relevance Labels}
\label{sec:appendix_relevance_label_strategy}

\begin{figure}[!t]
	\centering
	\includegraphics[width=0.95\linewidth]{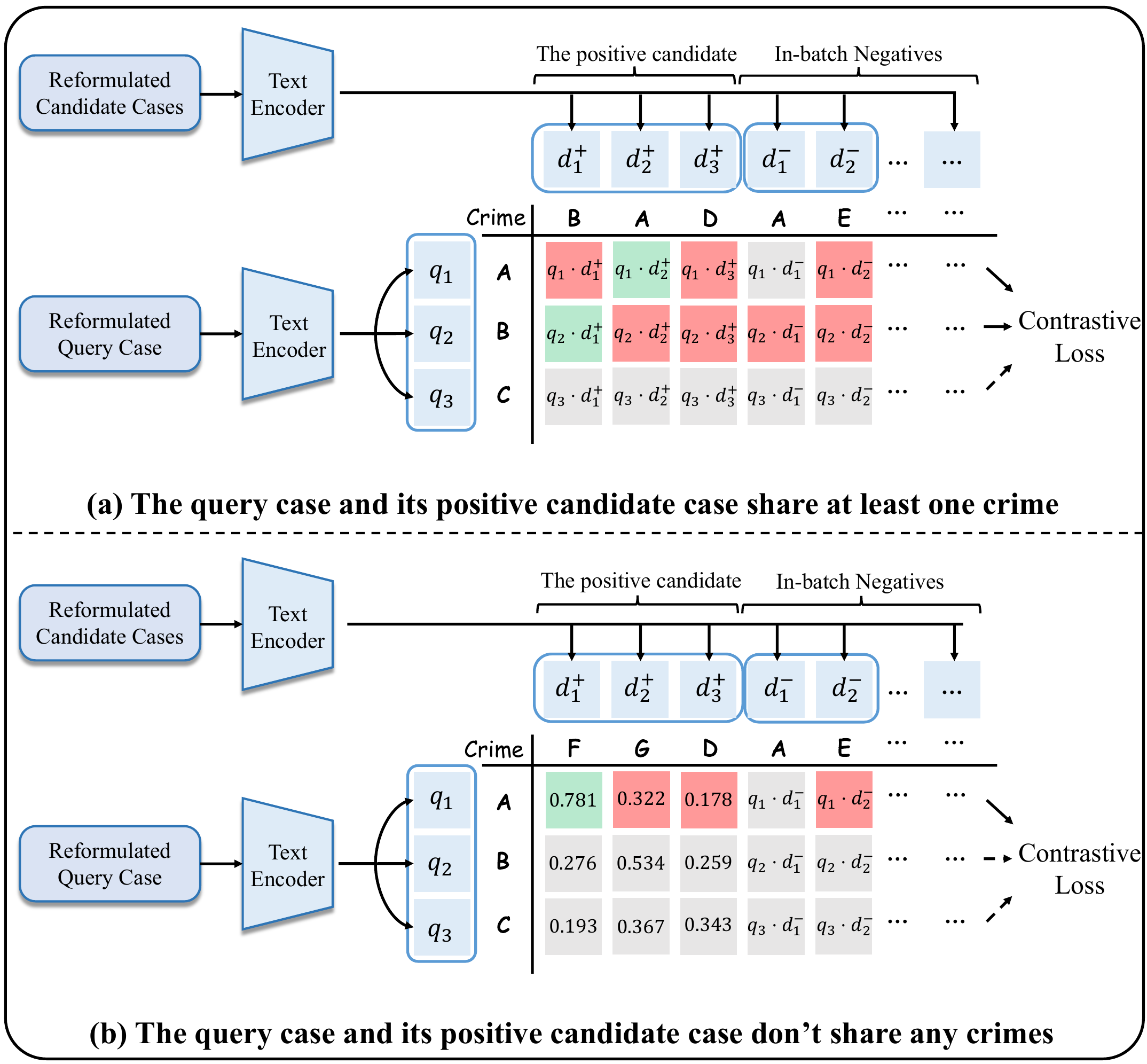}
	\caption{Illustration of our proposed sub-fact-level contrastive learning. The \textcolor{lightgreen}{green} and \textcolor{lightred}{red} squares represent the positive pairs and negative pairs, respectively. The \textcolor{gray}{gray} squares are the discarded pairs that are not used for training. The blue rounded rectangles encompass blue squares belonging to the same query/document case. \{A, ..., G\} are crimes.}
	\label{fig:Fig_CrimeCL}
\end{figure}

Specifically, \textbf{for a positive document} $d^+$ of query $q$, we first check whether any of the document sub-facts share the same crimes as any of the query sub-facts:
\begin{itemize}[leftmargin=*]
    \item If it exists, as shown in Figure~\ref{fig:Fig_CrimeCL}(a),  for a query sub-fact $q_i$, we treat the document sub-facts that share the same crime as the positives (e.g., the green rectangles in columns $d^{+}_{1}$, $d^{+}_{2}$, and $d^{+}_{3}$), and all the other document sub-facts as negatives (e.g., the red rectangles in columns $d^{+}_{1}$, $d^{+}_{2}$, and $d^{+}_{3}$). If the crime of $q_i$ is different from any of the document sub-facts, we will not include $q_i$ for training (e.g., the gray rectangles in row $q_3$).
    \item If not, as shown in Figure~\ref{fig:Fig_CrimeCL} (b), we select the $(q_i, d^{+}_j)$ which has the highest similarity score as a positive training pair (e.g., the green rectangle), and retain any $(q_i, d^{+}_k (k \neq j))$ as negatives (e.g., the red rectangles in columns $d^{+}_{2}$ and $d^{+}_{3}$). All the other query and document sub-fact pairs are discarded (e.g., the gray rectangles in columns $d^{+}_{1}$, $d^{+}_{2}$, and $d^{+}_{3}$). 
\end{itemize}

Then, \textbf{for a negative document} $d^-$ of one query sub-fact $q_i$, we first check whether $q_i$ has one positive sample.

\begin{itemize}[leftmargin=*]
    % \item If it exists, we will include all $(q_i, d_k^- (k \neq j))$ as negatives (e.g., the red rectangles of column $d^{-}_{1}$ and $d^{-}_{2}$ in Figure~\ref{fig:Fig_CrimeCL} (a) and (b)). All the other sub-facts are discarded to avoid introducing false negatives (e.g., the gray rectangles of column $d^{-}_{1}$ and $d^{-}_{2}$ in Figure~\ref{fig:Fig_CrimeCL} (a) and (b)). 
    % \item If not, we will include all ($q_i, d^-_j$) as negatives (e.g., the red rectangles of ($q_2, d^{-}_{1}$) and ($q_2, d^{-}_{2}$) in Figure~\ref{fig:Fig_CrimeCL} (a)).
    \item If not, we discard all the document sub-facts because there doesn't exist a positive sample for contrastive learning (e.g., the gray rectangles of row $q_3$ in Figure~\ref{fig:Fig_CrimeCL} (a) and (b)).
    \item If it exists, we further check whether one of its document sub-facts $d^-_j$ shares the same crime as a $q_i$. 
    \begin{enumerate}
        \item Both $d^-_j$ and $q_i$ are implicated to the same crime. we will include all $(q_i, d_k^- (k \neq j))$ as negatives (e.g., the red rectangles of column $d^{-}_{1}$ and $d^{-}_{2}$ in Figure~\ref{fig:Fig_CrimeCL} (a) and (b)). All the other sub-facts are discarded to avoid introducing false negatives (e.g., the gray rectangles of ($q_1, d^{-}_{1}$) in Figure~\ref{fig:Fig_CrimeCL} (a) and (b)).
        \item None of $d^-_j$ and $q_i$ pertain to the same crime. We will include all ($q_i, d^-_j$) as negatives (e.g., the red rectangles of ($q_2, d^{-}_{1}$) and ($q_2, d^{-}_{2}$) in Figure~\ref{fig:Fig_CrimeCL} (a)).
    \end{enumerate}
\end{itemize}

\section{Case Format of Other Regions}
\label{Appendix: case format}
\begin{figure}[!t]
	\centering
	\includegraphics[width=0.95\linewidth]{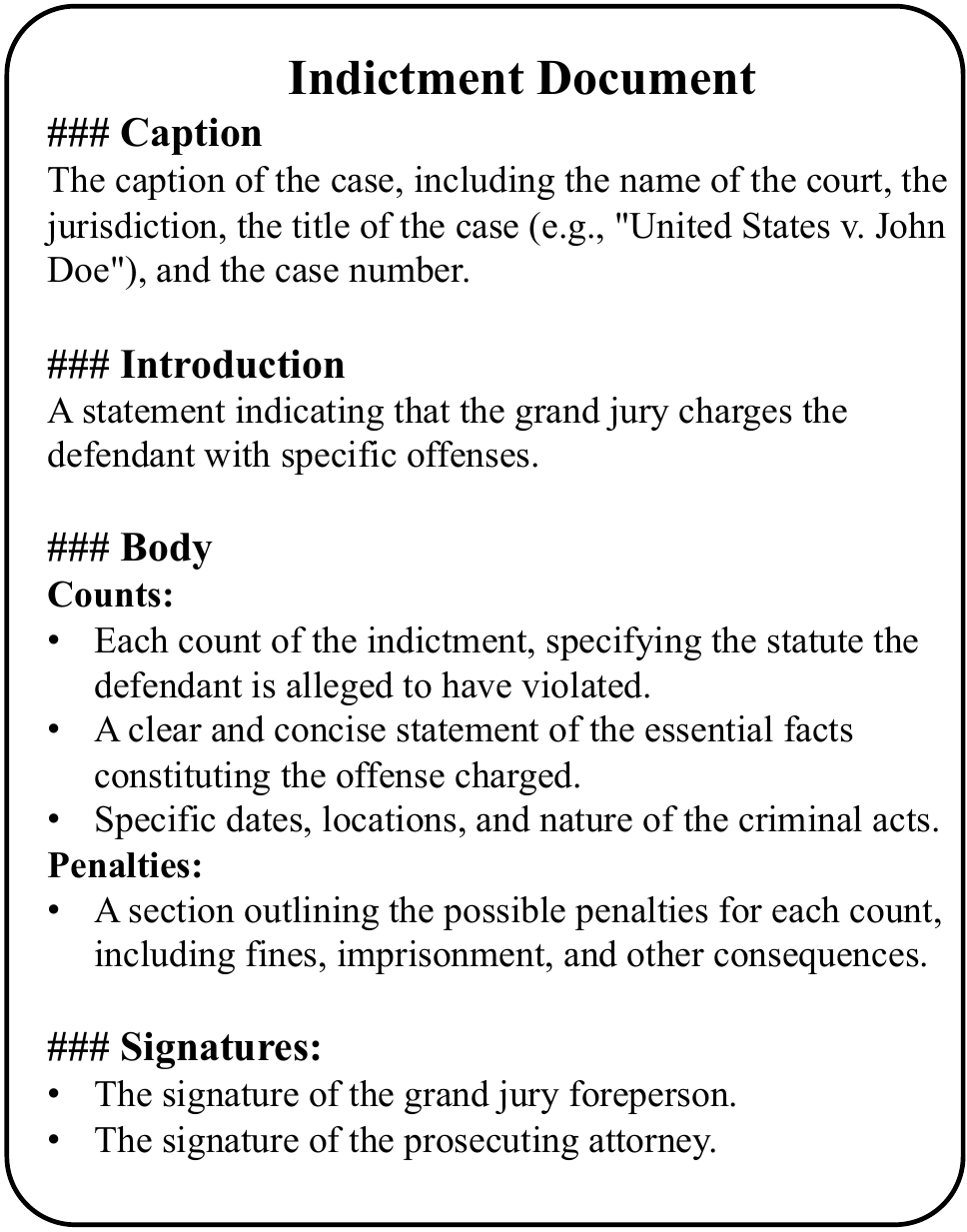}
	\caption{Illustration of the indictment document of US.}
	\label{fig:Fig_indictment}
\end{figure}

\begin{figure}[!t]
	\centering
	\includegraphics[width=0.95\linewidth]{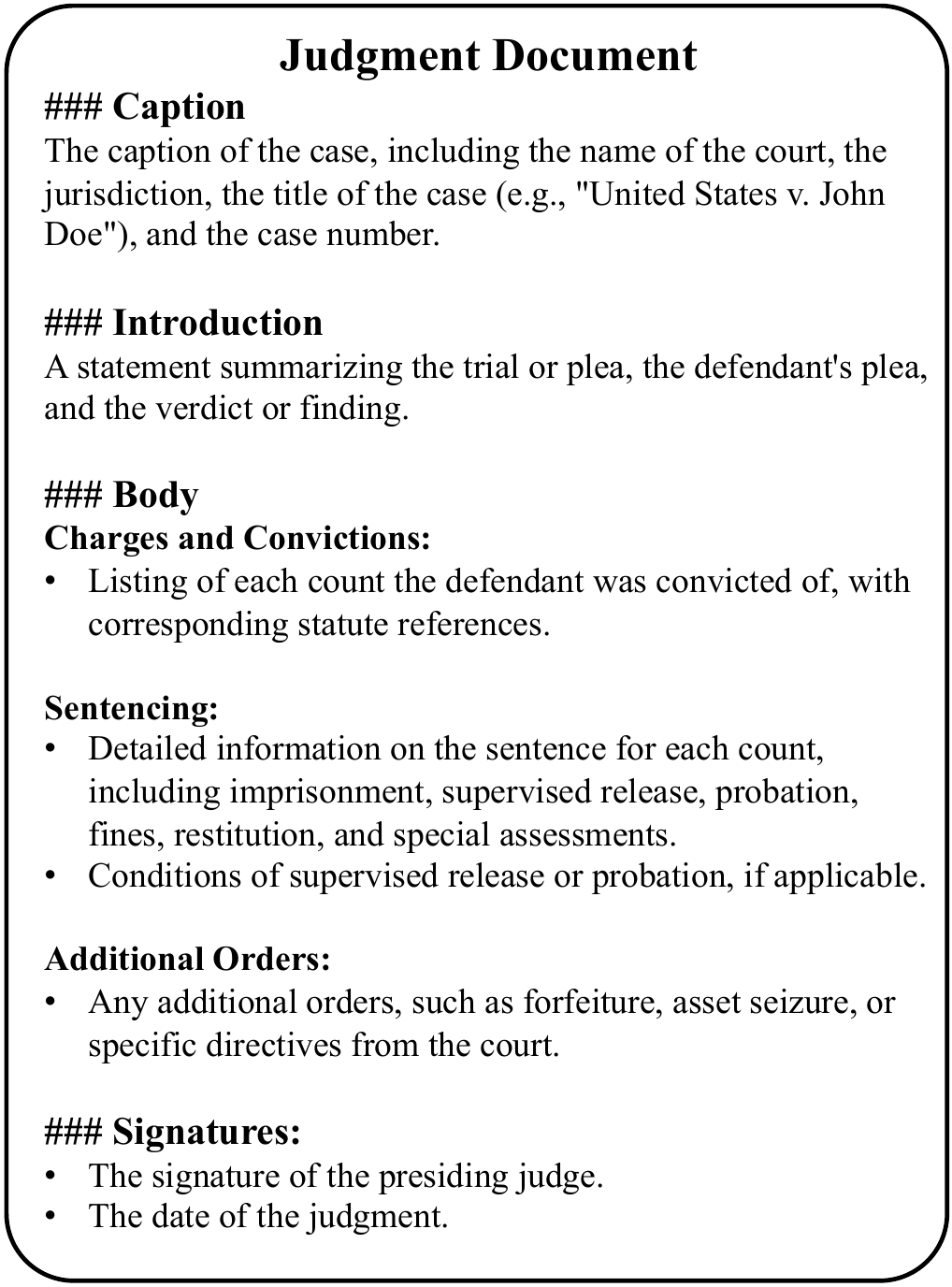}
	\caption{Illustration of the judgment document of US.}
	\label{fig:Fig_judgment}
\end{figure}

To demonstrate the international applicability of our method, we use U.S. legal documents as examples. Figure~\ref{fig:Fig_indictment} and Figure~\ref{fig:Fig_judgment} depict the formats of a U.S. indictment and a judgment document, respectively. It is evident that the legal knowledge required by our method (a combination of charges and law articles in this paper) is commonly present in the body sections of these documents. our method can be applied to reformulate legal texts in documents from other jurisdictions similarly, thereby enhancing their performance of legal case retrieval.


\begin{thebibliography}{32}
\providecommand{\natexlab}[1]{#1}

\bibitem[{Askari and Verberne(2021)}]{DBLP:conf/desires/AskariV21}
Arian Askari and Suzan Verberne. 2021.
\newblock \href {https://ceur-ws.org/Vol-2950/paper-02.pdf} {Combining lexical and neural retrieval with longformer-based summarization for effective case law retrieval}.
\newblock In \emph{Proceedings of the Second International Conference on Design of Experimental Search {\&} Information REtrieval Systems, Padova, Italy, September 15-18, 2021}, volume 2950 of \emph{{CEUR} Workshop Proceedings}, pages 162--170. CEUR-WS.org.

\bibitem[{Bai et~al.(2023)Bai, Bai, Chu, Cui, Dang, Deng, Fan, Ge, Han, Huang et~al.}]{bai2023qwen}
Jinze Bai, Shuai Bai, Yunfei Chu, Zeyu Cui, Kai Dang, Xiaodong Deng, Yang Fan, Wenbin Ge, Yu~Han, Fei Huang, et~al. 2023.
\newblock Qwen technical report.
\newblock \emph{arXiv preprint arXiv:2309.16609}.

\bibitem[{Chalkidis et~al.(2020)Chalkidis, Fergadiotis, Malakasiotis, Aletras, and Androutsopoulos}]{DBLP:journals/corr/abs-2010-02559}
Ilias Chalkidis, Manos Fergadiotis, Prodromos Malakasiotis, Nikolaos Aletras, and Ion Androutsopoulos. 2020.
\newblock \href {https://arxiv.org/abs/2010.02559} {{LEGAL-BERT:} the muppets straight out of law school}.
\newblock \emph{CoRR}, abs/2010.02559.

\bibitem[{Dai and Callan(2019)}]{dai2019deeper}
Zhuyun Dai and Jamie Callan. 2019.
\newblock Deeper text understanding for ir with contextual neural language modeling.
\newblock In \emph{Proceedings of the 42nd international ACM SIGIR conference on research and development in information retrieval}, pages 985--988.

\bibitem[{Devlin et~al.(2019)Devlin, Chang, Lee, and Toutanova}]{bert}
Jacob Devlin, Ming{-}Wei Chang, Kenton Lee, and Kristina Toutanova. 2019.
\newblock \href {https://doi.org/10.18653/V1/N19-1423} {{BERT:} pre-training of deep bidirectional transformers for language understanding}.
\newblock In \emph{Proceedings of the 2019 Conference of the North American Chapter of the Association for Computational Linguistics: Human Language Technologies, {NAACL-HLT} 2019, Minneapolis, MN, USA, June 2-7, 2019, Volume 1 (Long and Short Papers)}, pages 4171--4186. Association for Computational Linguistics.

\bibitem[{Gao et~al.(2023)Gao, Ma, Lin, and Callan}]{HyDE}
Luyu Gao, Xueguang Ma, Jimmy Lin, and Jamie Callan. 2023.
\newblock \href {https://doi.org/10.18653/V1/2023.ACL-LONG.99} {Precise zero-shot dense retrieval without relevance labels}.
\newblock In \emph{Proceedings of the 61st Annual Meeting of the Association for Computational Linguistics (Volume 1: Long Papers), {ACL} 2023, Toronto, Canada, July 9-14, 2023}, pages 1762--1777. Association for Computational Linguistics.

\bibitem[{Hamann(2019)}]{hamann2019german}
Hanjo Hamann. 2019.
\newblock The german federal courts dataset 1950--2019: from paper archives to linked open data.
\newblock \emph{Journal of empirical legal studies}, 16(3):671--688.

\bibitem[{Harris(2002)}]{harris2002final}
Bruce~V Harris. 2002.
\newblock Final appellate courts overruling their own" wrong" precedents: the ongoing search for principle.
\newblock \emph{Law Quarterly Review}, 118(July 2002):408--427.

\bibitem[{Jagerman et~al.(2023)Jagerman, Zhuang, Qin, Wang, and Bendersky}]{CoT_QueryRewriting}
Rolf Jagerman, Honglei Zhuang, Zhen Qin, Xuanhui Wang, and Michael Bendersky. 2023.
\newblock \href {https://doi.org/10.48550/ARXIV.2305.03653} {Query expansion by prompting large language models}.
\newblock \emph{CoRR}, abs/2305.03653.

\bibitem[{Khattab and Zaharia(2020)}]{khattab2020colbert}
Omar Khattab and Matei Zaharia. 2020.
\newblock Colbert: Efficient and effective passage search via contextualized late interaction over bert.
\newblock In \emph{Proceedings of the 43rd International ACM SIGIR conference on research and development in Information Retrieval}, pages 39--48.

\bibitem[{Lastres(2015)}]{lastres2015rebooting}
Steven~A Lastres. 2015.
\newblock Rebooting legal research in a digital age.

\bibitem[{Li et~al.(2023{\natexlab{a}})Li, Ai, Chen, Dong, Wu, Liu, Chen, and Tian}]{DBLP:conf/sigir/LiACDW0CT23}
Haitao Li, Qingyao Ai, Jia Chen, Qian Dong, Yueyue Wu, Yiqun Liu, Chong Chen, and Qi~Tian. 2023{\natexlab{a}}.
\newblock \href {https://doi.org/10.1145/3539618.3591761} {{SAILER:} structure-aware pre-trained language model for legal case retrieval}.
\newblock In \emph{Proceedings of the 46th International {ACM} {SIGIR} Conference on Research and Development in Information Retrieval, {SIGIR} 2023, Taipei, Taiwan, July 23-27, 2023}, pages 1035--1044. {ACM}.

\bibitem[{Li et~al.(2023{\natexlab{b}})Li, Shao, Wu, Ai, Ma, and Liu}]{li2023lecardv2}
Haitao Li, Yunqiu Shao, Yueyue Wu, Qingyao Ai, Yixiao Ma, and Yiqun Liu. 2023{\natexlab{b}}.
\newblock Lecardv2: A large-scale chinese legal case retrieval dataset.
\newblock \emph{arXiv preprint arXiv:2310.17609}.

\bibitem[{Liu et~al.(2019)Liu, Ott, Goyal, Du, Joshi, Chen, Levy, Lewis, Zettlemoyer, and Stoyanov}]{DBLP:journals/corr/abs-1907-11692}
Yinhan Liu, Myle Ott, Naman Goyal, Jingfei Du, Mandar Joshi, Danqi Chen, Omer Levy, Mike Lewis, Luke Zettlemoyer, and Veselin Stoyanov. 2019.
\newblock \href {https://arxiv.org/abs/1907.11692} {Roberta: {A} robustly optimized {BERT} pretraining approach}.
\newblock \emph{CoRR}, abs/1907.11692.

\bibitem[{Ma et~al.(2023{\natexlab{a}})Ma, Gong, He, Zhao, and Duan}]{Query_Rewriting_for_Retrieval-Augmented_Large_Language_Models}
Xinbei Ma, Yeyun Gong, Pengcheng He, Hai Zhao, and Nan Duan. 2023{\natexlab{a}}.
\newblock \href {https://doi.org/10.48550/ARXIV.2305.14283} {Query rewriting for retrieval-augmented large language models}.
\newblock \emph{CoRR}, abs/2305.14283.

\bibitem[{Ma et~al.(2021)Ma, Shao, Wu, Liu, Zhang, Zhang, and Ma}]{ma2021lecard}
Yixiao Ma, Yunqiu Shao, Yueyue Wu, Yiqun Liu, Ruizhe Zhang, Min Zhang, and Shaoping Ma. 2021.
\newblock Lecard: a legal case retrieval dataset for chinese law system.
\newblock In \emph{Proceedings of the 44th international ACM SIGIR conference on research and development in information retrieval}, pages 2342--2348.

\bibitem[{Ma et~al.(2023{\natexlab{b}})Ma, Wu, Su, Ai, and Liu}]{DBLP:conf/emnlp/MaWSA023}
Yixiao Ma, Yueyue Wu, Weihang Su, Qingyao Ai, and Yiqun Liu. 2023{\natexlab{b}}.
\newblock \href {https://aclanthology.org/2023.emnlp-main.441} {Caseencoder: {A} knowledge-enhanced pre-trained model for legal case encoding}.
\newblock In \emph{Proceedings of the 2023 Conference on Empirical Methods in Natural Language Processing, {EMNLP} 2023, Singapore, December 6-10, 2023}, pages 7134--7143. Association for Computational Linguistics.

\bibitem[{Mao et~al.(2023)Mao, Dou, Mo, Hou, Chen, and Qian}]{llm4cs}
Kelong Mao, Zhicheng Dou, Fengran Mo, Jiewen Hou, Haonan Chen, and Hongjin Qian. 2023.
\newblock \href {https://aclanthology.org/2023.findings-emnlp.86} {Large language models know your contextual search intent: {A} prompting framework for conversational search}.
\newblock In \emph{Findings of the Association for Computational Linguistics: {EMNLP} 2023, Singapore, December 6-10, 2023}, pages 1211--1225. Association for Computational Linguistics.

\bibitem[{Saravanan et~al.(2009)Saravanan, Ravindran, and Raman}]{saravanan2009improving}
Manavalan Saravanan, Balaraman Ravindran, and Shivani Raman. 2009.
\newblock Improving legal information retrieval using an ontological framework.
\newblock \emph{Artificial Intelligence and Law}, 17:101--124.

\bibitem[{Shao et~al.(2020)Shao, Mao, Liu, Ma, Satoh, Zhang, and Ma}]{shao2020bert}
Yunqiu Shao, Jiaxin Mao, Yiqun Liu, Weizhi Ma, Ken Satoh, Min Zhang, and Shaoping Ma. 2020.
\newblock Bert-pli: Modeling paragraph-level interactions for legal case retrieval.
\newblock In \emph{IJCAI}, pages 3501--3507.

\bibitem[{Tang et~al.(2023)Tang, Qiu, and Li}]{DBLP:conf/adc/TangQL23}
Yanran Tang, Ruihong Qiu, and Xue Li. 2023.
\newblock \href {https://doi.org/10.1007/978-3-031-47843-7\_7} {Prompt-based effective input reformulation for legal case retrieval}.
\newblock In \emph{Databases Theory and Applications - 34th Australasian Database Conference, {ADC} 2023, Melbourne, VIC, Australia, November 1-3, 2023, Proceedings}, volume 14386 of \emph{Lecture Notes in Computer Science}, pages 87--100. Springer.

\bibitem[{Tran et~al.(2020)Tran, Le~Nguyen, Tojo, and Satoh}]{tran2020encoded}
Vu~Tran, Minh Le~Nguyen, Satoshi Tojo, and Ken Satoh. 2020.
\newblock Encoded summarization: summarizing documents into continuous vector space for legal case retrieval.
\newblock \emph{Artificial Intelligence and Law}, 28:441--467.

\bibitem[{Wang et~al.(2023)Wang, Yang, and Wei}]{Query2Doc}
Liang Wang, Nan Yang, and Furu Wei. 2023.
\newblock \href {https://aclanthology.org/2023.emnlp-main.585} {Query2doc: Query expansion with large language models}.
\newblock In \emph{Proceedings of the 2023 Conference on Empirical Methods in Natural Language Processing, {EMNLP} 2023, Singapore, December 6-10, 2023}, pages 9414--9423. Association for Computational Linguistics.

\bibitem[{Xiao et~al.(2021)Xiao, Hu, Liu, Tu, and Sun}]{xiao2021lawformer}
Chaojun Xiao, Xueyu Hu, Zhiyuan Liu, Cunchao Tu, and Maosong Sun. 2021.
\newblock Lawformer: A pre-trained language model for chinese legal long documents.
\newblock \emph{AI Open}, 2:79--84.

\bibitem[{Xiao et~al.(2023)Xiao, Liu, Zhang, and Muennighof}]{DBLP:journals/corr/abs-2309-07597}
Shitao Xiao, Zheng Liu, Peitian Zhang, and Niklas Muennighof. 2023.
\newblock \href {https://doi.org/10.48550/ARXIV.2309.07597} {C-pack: Packaged resources to advance general chinese embedding}.
\newblock \emph{CoRR}, abs/2309.07597.

\bibitem[{Xiong et~al.(2017)Xiong, Dai, Callan, Liu, and Power}]{KNRM}
Chenyan Xiong, Zhuyun Dai, Jamie Callan, Zhiyuan Liu, and Russell Power. 2017.
\newblock \href {https://doi.org/10.1145/3077136.3080809} {End-to-end neural ad-hoc ranking with kernel pooling}.
\newblock In \emph{Proceedings of the 40th International {ACM} {SIGIR} Conference on Research and Development in Information Retrieval, Shinjuku, Tokyo, Japan, August 7-11, 2017}, pages 55--64. {ACM}.

\bibitem[{Yao et~al.(2022)Yao, Xiao, Wang, Liu, Hou, Tu, Li, Liu, Shen, and Sun}]{DBLP:conf/acl/YaoXWL0TLLSS22}
Feng Yao, Chaojun Xiao, Xiaozhi Wang, Zhiyuan Liu, Lei Hou, Cunchao Tu, Juanzi Li, Yun Liu, Weixing Shen, and Maosong Sun. 2022.
\newblock \href {https://doi.org/10.18653/v1/2022.findings-acl.17} {{LEVEN:} {A} large-scale chinese legal event detection dataset}.
\newblock In \emph{Findings of the Association for Computational Linguistics: {ACL} 2022, Dublin, Ireland, May 22-27, 2022}, pages 183--201. Association for Computational Linguistics.

\bibitem[{Yu et~al.(2022)Yu, Sun, Xu, Dong, Chen, Xu, and Wen}]{DBLP:conf/sigir/YuS0DCXW22}
Weijie Yu, Zhongxiang Sun, Jun Xu, Zhenhua Dong, Xu~Chen, Hongteng Xu, and Ji{-}Rong Wen. 2022.
\newblock \href {https://doi.org/10.1145/3477495.3531974} {Explainable legal case matching via inverse optimal transport-based rationale extraction}.
\newblock In \emph{{SIGIR} '22: The 45th International {ACM} {SIGIR} Conference on Research and Development in Information Retrieval, Madrid, Spain, July 11 - 15, 2022}, pages 657--668. {ACM}.

\bibitem[{Zeng et~al.(2005)Zeng, Wang, Zeleznikow, and Kemp}]{DBLP:conf/kes/ZengWZK05}
Yiming Zeng, Ruili Wang, John Zeleznikow, and Elizabeth~A. Kemp. 2005.
\newblock \href {https://doi.org/10.1007/11552413\_49} {Knowledge representation for the intelligent legal case retrieval}.
\newblock In \emph{Knowledge-Based Intelligent Information and Engineering Systems, 9th International Conference, {KES} 2005, Melbourne, Australia, September 14-16, 2005, Proceedings, Part {I}}, volume 3681 of \emph{Lecture Notes in Computer Science}, pages 339--345. Springer.

\bibitem[{Zhang et~al.(2023)Zhang, Chen, Wang, Tian, and Bai}]{zhang2023cfgl}
Kun Zhang, Chong Chen, Yuanzhuo Wang, Qi~Tian, and Long Bai. 2023.
\newblock Cfgl-lcr: A counterfactual graph learning framework for legal case retrieval.
\newblock In \emph{Proceedings of the 29th ACM SIGKDD Conference on Knowledge Discovery and Data Mining}, pages 3332--3341.

\bibitem[{Zhong et~al.(2019)Zhong, Zhang, Liu, and Sun}]{zhong2019openclap}
Haoxi Zhong, Zhengyan Zhang, Zhiyuan Liu, and Maosong Sun. 2019.
\newblock \href {https://github.com/thunlp/openclap} {Open chinese language pre-trained model zoo}.
\newblock Technical report.

\bibitem[{Zhu et~al.(2023)Zhu, Yuan, Wang, Liu, Liu, Deng, Dou, and Wen}]{LLM4IR_Survey}
Yutao Zhu, Huaying Yuan, Shuting Wang, Jiongnan Liu, Wenhan Liu, Chenlong Deng, Zhicheng Dou, and Ji{-}Rong Wen. 2023.
\newblock \href {https://doi.org/10.48550/ARXIV.2308.07107} {Large language models for information retrieval: {A} survey}.
\newblock \emph{CoRR}, abs/2308.07107.

\end{thebibliography}
\end{document}